\documentclass[conference, a4paper]{IEEEtran}
\IEEEoverridecommandlockouts

	\usepackage[turkish]{babel}
	\usepackage[utf8]{inputenc} 
	\usepackage[T1]{fontenc}
\usepackage{amssymb}

\usepackage{cite}

\ifCLASSINFOpdf
  \usepackage[pdftex]{graphicx}

\else
\fi
\usepackage[cmex10]{amsmath}
\usepackage{multirow}
\usepackage{array}
\usepackage[lofdepth,lotdepth]{subfig}
\usepackage{xfrac}

\AtBeginDocument{%
  \renewcommand\tablename{TABLO}
}
\AtBeginDocument{%
  \renewcommand\abstractname{Abstract}
}

\begin{document}
\bstctlcite{IEEEexample:BSTcontrol}
\IEEEpubid{\makebox[\columnwidth]{978-1-7281-7206-4/20/\$31.00 ©2020 IEEE\hfill}
\hspace{\columnsep}\makebox[\columnwidth]{}}

%
\title{Çöz-Aktar İşbirlikli NOMA Sistemlerinin Nakagami-m Sönümlemeli Kanallarda Hata Analizi \\
Error Analysis of Decode-Forward Cooperative Relaying NOMA Schemes over Nakagami-m Fading Channels }
\author{\IEEEauthorblockN{
Ferdi KARA, Hakan KAYA}
\IEEEauthorblockA{
Kablosuz İletişim Teknolojileri Araştırma Laboratuvarı (KİTLab) \\
Elektrik-Elektronik Mühendisliği\\
Zonguldak Bülent Ecevit Üniversitesi\\
Zonguldak, TÜRKİYE 67100\\
E-posta: \{f.kara,hakan.kaya\}@beun.edu.tr}
}


%

\maketitle

\begin{ozet}
Dik-olmayan çoklu erişim (NOMA) sağladığı yüksek spektral verimlilik sayesinde son yıllarda pek çok araştırmaya konu olmuştur. En fazla araştırılan konuların başında gelen işbirlikli NOMA sistemleri de hem akademiden hem de endüstriden bir çok araştırmacının oldukça ilgisini çekmiştir. Fakat, yapılan çalışmaların çoğunluğunda, işbirlikli NOMA yalnızca ulaşılabilir kapasite ve kesinti olasılığı açısından incelenirken, en önemli başarım kriterlerinden biri olan bit hata olasılığı (BHO) göz ardı edilmiştir. Bu çalışmada, çöz-aktar protokü kullanılan işbirlikli NOMA sistemlerinin BHO başarımları incelenmiştir. Daha genel bir kanal modeli olan Nakagami-m sönümleli kanallar için BHO ifadeleri kapalı formda elde edilmiştir. Türetilen ifadeler bilgisayar benzetimleri ile doğrulanmıştır.
\end{ozet}
\begin{IEEEanahtar}
işbirlikli iletişim, dik-olmayan çoklu erişim, hata analizi, Nakagami-m
\end{IEEEanahtar}

\begin{abstract}
Non-orthogonal multiple access (NOMA) has attracted great recent attention due to its high spectral efficiency. Besides, NOMA can be easily implemented in physical layer, thus its interplays with other psychical layer techniques have been analyzed widely by researches. Cooperative NOMA schemes are the most investigated among these interplays. However, these studies mostly analyze cooperative NOMA schemes in terms of achievable rate and outage probability whereas bit error probability (BEP) for those systems has not been studied well although it is one of the most important performance metrics. In this paper, we investigate the error performance of cooperative NOMA schemes where a decode-forward relay helps NOMA users. We derive exact BEP expressions in closed-form over Nakagami-m fading channels. All derived expressions are validated via computer simulations.
\end{abstract}



%


\section{G{\footnotesize İ}r{\footnotesize İ}ş}

Dik-olmayan çoklu erişim (non-orthogonal multiple access -NOMA) sağladığı yüksek verimlilik sayesinde 5G ve sonrası kablosuz ağlar için en önemli adaylardan biri konumundadır. NOMA farklı kullanıcıların aynı radyo kaynağını farklı güç paylaşım katsayıları ile kullanılması prensbine dayanmaktadır. Bu işlem süperpozisyon kodlaması olarak adlandırılır. Aynı radyo kaynaklarının kullanılması sonucunda oluşan kullanıcılar arası girişim alıcılırda ardışık girişim giderici (successive interference canceler -SIC) kullanarak giderilir \cite{Saito2013}. NOMA'nın fiziksel seviye açısından bu kadar kolay uygulanabilir olması, NOMA ve diğer fiziksel seviye tekniklerin beraber uygulandığı çalışmaların hızla artmasına sebep olmuştur. Bunların başında da işbirlikli iletişim ve NOMA'nın beraber uygulandığı çalışmalar gelemektedir \cite{Kim2015a,Liu2015b}. \cite{Liu2016d}'te yazarlar NOMA kullanıcılarına bir röle yardımıyla iletişimin sağlandığı sistemi ele almışlar ve ulaşılabilir kapasite/hız analizleri sunulmuştur. \cite{Liu2016d}'te verilen analizler farklı kanal modelleri için genişletilmiştir \cite{Liu2018b}. Daha sonra ortamda bulunan rölenin iki farklı NOMA ağı tarafınfan paylaşıldığı durumda kesinti olasılığı analizleri sunulmuştur \cite{Kader2017}. Kaynak ile NOMA kullanıcıları arasında direkt yolun da bulunduğu durum için kapasite ve kesinti olasılığı analizleri de verilmiştir \cite{Liu2018}. Ayrıca, ortamda birden çok rölenin bulunduğu durumda, NOMA kullanıcıları için röle seçimi algoritmaları da araştırılmış ve önerilen algoritmalar için kapasite ile kesinti olasılığı ifadeleri türetilmiştir \cite{Ding2016c,Li2019a}. Fakat, yukarıda da belirtildiği gibi işbirlikli NOMA çalışmaları çoğunlukla yalnızca kapasite ve kesinti olasılığı açısından araştırılmış olup hata performansları göz ardı edilmiştir. Sadece \cite{Karaa}'da Rayleigh sönümlemeli kanallar için bit hata olasığı ifadeleri sunulmuştur. Fakat, daha genel kanal modelleri için literatürde herhangi bir çalışma yapılmamıştır.

Bu çalışmada, çöz-aktar protokolü kullanan bir işbirlikli NOMA sistemi için bit hata olasılığı (BHO) ifadeleri Nakagami-m sönümlemeli kanallarda kapalı formda türetilmiştir. Elde edilen ifadeler Monte Carlo benzetimleri ile doğrulanmıştır. Çalışmanın bundan sonraki bölümleri aşağıdaki gibidir. Bölüm II'de çöz-aktar işbirlikli NOMA sistemi tanıtılmaktadır. Bölüm III'te, BHO ifadelerinin teorik çıkarımları verilmiştir. Daha sonra Bölüm IV'te, elde edilen ifadelerin doğrulamaları bilgisayar benzetimleri ile sunulmuştur. Son olarak, Bölüm V'te sonuçlar tartışılmış ve çalışma sonlandırılmıştır.

\section{S{\footnotesize İ}stem Model{\footnotesize İ}}
Tüm düğümlerin tek antene sahip olduğu bir işbirlikli iletişim sistemi ele alınmıştır. Sistem modelinde bir kaynak (S), çöz-aktar protokolü kullanan bir röle (R) ve iki adet kulanıcı ($D_i$, $i=1,2$) bulunmaktadır. Yarı çift yönlü iletişim düşünülmektedir. Bu nedenle toplam iletişim iki zaman diliminde tamamlanmaktadır. Sistem modeli Şekil 1'de verilmiştir.
\begin{figure}[!t]
	\centering
	\shorthandoff{=}  
	\includegraphics[width=8.5cm, height=3cm]{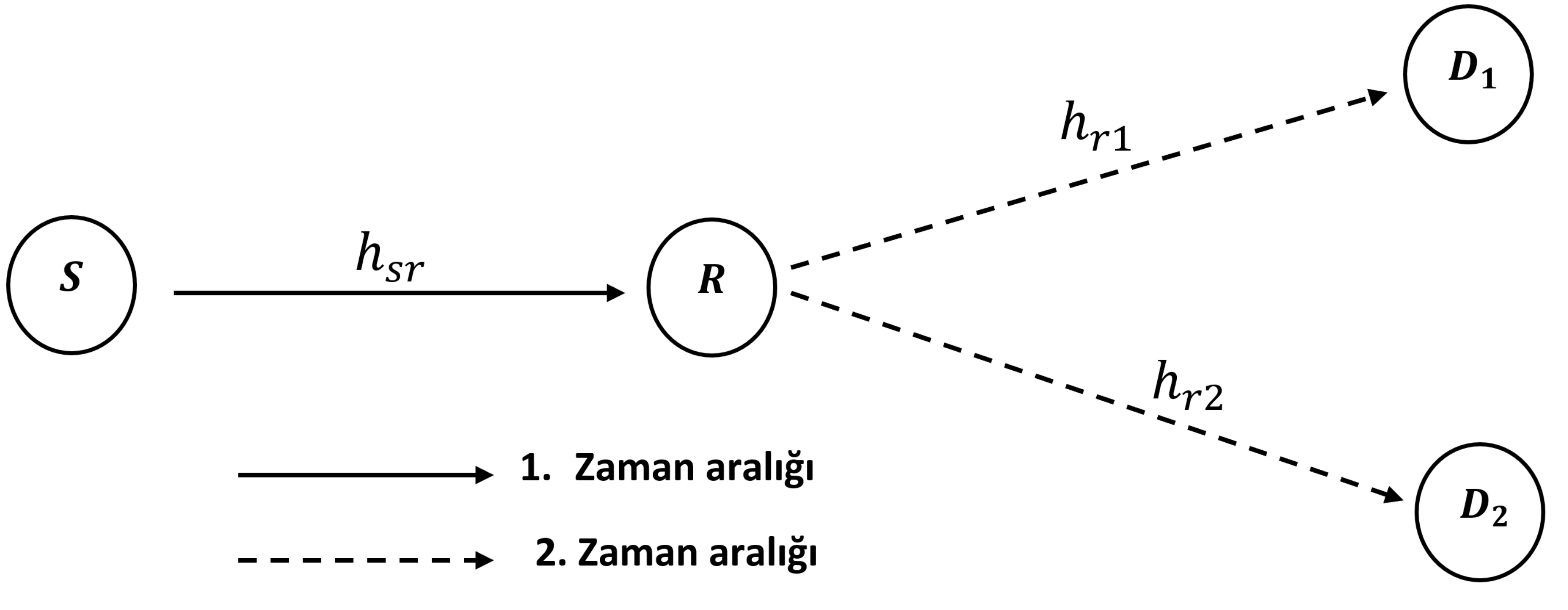}
	\shorthandon{=} 
	\caption{Çöz-Aktar İşbirlikli NOMA Sistemi}
	\label{sekil1}
\end{figure}
Spektral verimiliği artırmak adına kullanıcıların sembolleri NOMA kullanılarak iletildiğinden, birinci zaman diliminde kaynaktan gönderilen toplam işaret, rölede
\begin{equation}
  y_{sr}=\sqrt{P_s}\left(\sqrt{\alpha_1}x_1+\sqrt{\alpha_2}x_2\right)h_{sr}+n_{sr}
\end{equation}
olarak alınır. Burada $P_s$ kaynak iletim gücünü, $h_{sr}$ kaynak-röle arasındaki sönümlemeli kanal katsayısını ve $n_{sr}$ toplamsal Gauss gürültüsünü göstermektedir. $h_{sr}$'nin zarfının Nakagami-m dağılımlı olduğu varsayılmaktadır. $n_{sr}$ sıfır beklenen değer ve $N_0$ varsanya sahiptir. Denklem (1)'de verilen $\sqrt{\alpha_i}$, $x_i$, $i=1,2$ çiftleri sırasıyla kullanıcı $D_i$ için güç paylaşım katsayısını ve kullanıcıların temel bant modulasyonlu sembollerini göstermektedir. Burada $\alpha_2>\alpha_1$ ve $\alpha_1+\alpha_2=1$ olarak verilir. $x_i$, $i=1,2$ sembollerinin ikili faz kaydırmalı anahtarlama kullanılarak (binary phase shift keying -BPSK) modüle edildiği varsayılmıştır. Alınan $y_{sr}$ işaretine göre, röle öncelikle $x_1$ sembollerini gürültü gibi düşünerek $x_2$ sembollerini demodüle eder/çözer. Daha sonra demodüle edilen sembolleri tekrar modüle ederek $\hat{x}_2$ sembollerini alınan $y_{sr}$'den çıkarır ve $x_1$ sembollerini demodüle eder/çözer. Bu işlem SIC olarak adlandırılır. Röle her iki kullanıcının sembollerini elde ettikten sonra, ikinci zaman diliminde tekrar süperpozisyon uygulayarak kullancılara gönderir. İkinci zaman diliminde her bir kullanıcıda alınan işaret
\begin{equation}
  y_{ri}=\sqrt{P_r}\left(\sqrt{\beta_1}\hat{x}_1+\sqrt{\beta_2}\hat{x}_2\right)h_{ri}+n_{ri}, \ i=1,2
\end{equation}
olarak verilir. Burada $P_r$ rölenin iletim gücünü, $h_{ri}$ röle ile $i.$ kullanıcı arasındaki sönümleme katsayısını göstermektedir. $\hat{x}_i$ ve $\beta_i$, $i=1,2$ sırasıyla rölede demodüle-tekrar modüle edilen $i.$ kullanıcının temel bant işaretini ve $i.$ kullanıcı için güç paylaşım katsayısını göstermektedir. $\beta_2>\beta_1$ ve $\beta_1+\beta_2=1$ olarak verilir. Son olarak, kullanıcılar alınan $y_{ri}$ işaretine göre kendi sembollerini kestitirler. Bu aşamada $D_2$ kullanıcısı $\hat{x}_1$ sembollerine gürültü gibi davranarak kendi sembollerini demodüle ederken, $D_1$ kullanıcı SIC işlemi uygulamalıdır.



\section{B{\footnotesize İ}t Hata Olasılığı Anal{\footnotesize İ}z{\footnotesize İ}}
Çöz-aktar protokolü kullanan bir röle kullanıldığından sistemin uçtan-uca (end-to-end -e2e) toplam BHO'sunu bulabilmek için her zaman dilimindeki BHO ifadeleri bulunmalıdır. Her iki zaman dilimindeki hata olaylarının istatiksel olarak bağımsız olduğu düşünülürse, her bir kullanıcı için e2e BHO ifadesi, toplam olasılık kuralından yararlanılarak
\begin{equation}
\begin{split}
  P_i^{(e2e)}(e)=&P_i^{(sr)}(e)\times\left(1-P_i^{(ri)}(e)\right)\\
  &+\left(1-P_i^{(sr)}(e)\right)P_i^{(ri)}(e), i=1,2
\end{split}
\end{equation}
olarak bulunur. Burada $P_i^{(sr)}(e)$ birinci zaman diliminde kaynak ile röle arasındaki hata olasılığını ve $P_i^{(ri)}(e)$ de ikinci zaman diliminde röle ile $i.$ kullanıcı $D_i$ arasındaki hata olasılığını göstermektedir. İlk olarak kaynak ve röle arasndaki iletişimi ele alalım. Bir aşağı yönlü NOMA sistemi gibi düşünülebilir ve röledeki hata olasılıkları geleneksel NOMA'daki yakın kullanıcı hata olasıklarına dönüşür. NOMA'da süperpozisyon uygulandığı için toplam sembolün işaret-yıldız kümesi de süperpozisyona uğramış olur. Bu durumda demodüle/çözme sırasında öncelikli olan $x_2$ sembolleri için hata olasılığı ifadesi bu toplam işaret-yıldız kümsesine göre bulunur. Her iki kullanıcının da BPSK ile modüle edildiği durumda $x_2$ sembolleri için koşullu hata olasılığı ifadesi
\begin{equation}
  P_2^{(sr)}(e|_{\gamma_{sr}})=\sum_{k=1}^{2}\sigma_kQ\left(\sqrt{2\lambda_{k}\rho_s\gamma_{sr}}\right)
\end{equation}
olarak verilir \cite{Kara2019a}. Burada $\gamma_{sr}\triangleq\left|h_{sr}\right|^2$, $\rho_s=\sfrac{P_s}{N_0}$, $\sigma_k=0.5, k=1,2$ ve $\lambda_k=\left(\sqrt{\alpha_2}\mp\sqrt{\alpha_1}\right)^2$, $k=1,2$ şeklinde tanımlanır. $h_{sr}$ Nakagami-m dağılımlı olduğu durumda $\gamma_{sr}$ Gamma dağılımına sahip olur. Ortalama hata olasılığı ifadesini bulabilmek için (4)'te verilen ifadenin Gamma dağılımının olasılık yoğunluk fonksiyonu kullanılarak integrali alınırsa, \cite{Simon2004} yardımıyla
\begin{equation}
\begin{split}
  &P_2^{(sr)}(e)=\\
  &\begin{cases}
  \sum\limits_{k=1}^2\frac{\sigma_k}{2}\left[1-\mu^2(a_k)\sum\limits_{l=0}^{m_{sr}-1}\binom{2l}{l}\left(\frac{1-\mu^2(a_k)}{4}\right)^l\right], \\
  \quad \quad \quad \quad \quad \quad \quad \quad \quad \quad \quad \quad \quad \quad \quad \quad \quad \quad \quad m_{sr} \ \text{tamsayı} \\
  \sum\limits_{k=1}^2\frac{\sigma_k}{2\sqrt{\pi}}\frac{\sqrt{a_k}}{\left(1+a_k\right)^{m_{sr}+0.5}}\frac{\Gamma(m_{sr}+0.5)}{\Gamma(m_{sr}+1)} \times\\
  {_2}F_1\left(1,m_{sr}+0.5;m_{sr}+1;\frac{1}{1+a_k}\right), \quad   m_{sr} \ \text{tamsayı değil}
    \end{cases}
    \end{split}
 \end{equation}
olarak bulunur. Burada,  Burada $a_k=\frac{\lambda_k\rho_s\Omega_{sr}}{2m_{sr}}$, $\Omega_{sr}=E[\gamma_{sr}]$ olarak tanımlanır. $\mu(z)=\sqrt{\frac{z}{1+z}}$ olarak verilir \cite{Kara2018c}. $\Gamma(.)$ ve $ {_2}F_1\left(:,:;:;:\right)$ sırasıyla Gamma \cite[Eq. (8.31)]{Gradshteyn1994} ve Gauss Hipergeometrik \cite[Eq. (9.10)]{Gradshteyn1994} fonksiyonlarıdır.

$x_1$ sembollerinin hata olasılığı gerçekleştirilen SIC işleminin başarımına bağlıdır. Bu nedenle doğru ve hatalı SIC işlemlerinin gerçekleştiği koşullar ayrı ayrı değerlendirilmelidir. Kaynak röle arasındaki koşullu hata olasığını bulabilmek için \cite[Eq. (6)- Eq. (9)]{Kara2019} adımları tekrarlanırsa,
\begin{equation}
  P_1^{(sr)}(e|_{\gamma_{sr}})=\sum_{k=1}^{5}\tau_k Q\left(\sqrt{2\nu_{k}\rho_s\gamma_{sr}}\right)
\end{equation}
olarak bulunur. Burada, $\tau_k=\sfrac{1}{2}[-1, 1, 2, 1, -1]$ ve \\ $\nu_k=\left[\left(\sqrt{\alpha_2}+\sqrt{\alpha_1}\right)^2, \left(\sqrt{\alpha_2}-\sqrt{\alpha_1}\right)^2, \alpha_1, \left(2\sqrt{\alpha_2}+\sqrt{\alpha_1}\right)^2, \right. \\ \left. \left(2\sqrt{\alpha_2}-\sqrt{\alpha_1}\right)^2\right]$ olarak verilir. Denklem (6)'da verilen koşullu olasılık ifadesinin ortalaması alınırsa, (5)'te olduğu gibi
\begin{equation}
\begin{split}
  &P_1^{(sr)}(e)=\\
  &\begin{cases}
  \sum\limits_{k=1}^5\frac{\tau_k}{2}\left[1-\mu^2(b_k)\sum\limits_{l=0}^{m_{sr}-1}\binom{2l}{l}\left(\frac{1-\mu^2(b_k)}{4}\right)^l\right], \\
  \quad \quad \quad \quad \quad \quad \quad \quad \quad \quad \quad \quad \quad \quad \quad \quad \quad \quad \quad m_{sr} \ \text{tamsayı} \\
  \sum\limits_{k=1}^5\frac{\tau_k}{2\sqrt{\pi}}\frac{\sqrt{b_k}}{\left(1+b_k\right)^{m_{sr}+0.5}}\frac{\Gamma(m_{sr}+0.5)}{\Gamma(m_{sr}+1)} \times\\
  {_2}F_1\left(1,m_{sr}+0.5;m_{sr}+1;\frac{1}{1+b_k}\right), \quad   m_{sr} \ \text{tamsayı değil}
    \end{cases}
    \end{split}
 \end{equation}
şeklinde elde edilir. Burada $b_k=\frac{\nu_k\rho_s\Omega_{sr}}{2m_{sr}}$ olarak tanımlanır.

Rölede demodüle edilen semboller tekrar modüle edilip farklı güç katsayıları ile kullanıcılara gönderildiğinden, ikinci zaman dilimindeki iletişim geleneksel aşağı yönlü NOMA iletişimine dönüşmektedir. Dolayısyla ikinci zaman dilimi için hata olasılığı ifadeleri, birinci zaman diliminde yapılan işlemler tekrarlanarak elde edilebilir. Uzak kullanıcı olarak adlandırılan $D_2$ kendi sembollerini $\hat{x}_1$ sembollerine gürültü gibi davranarak demodüle ettiğinden, $R-D_2$ arasındaki ortalama hata olasılığı ifadesi (4)-(5) yardımıyla
 \begin{equation}
\begin{split}
  &P_2^{(r2)}(e)=\\
  &\begin{cases}
  \sum\limits_{k=1}^2\frac{\sigma_k}{2}\left[1-\mu^2(c_k)\sum\limits_{l=0}^{m_{r2}-1}\binom{2l}{l}\left(\frac{1-\mu^2(c_k)}{4}\right)^l\right], \\
  \quad \quad \quad \quad \quad \quad \quad \quad \quad \quad \quad \quad \quad \quad \quad \quad \quad \quad \quad m_{r2} \ \text{tamsayı} \\
  \sum\limits_{k=1}^2\frac{\sigma_k}{2\sqrt{\pi}}\frac{\sqrt{c_k}}{\left(1+c_k\right)^{m_{r2}+0.5}}\frac{\Gamma(m_{r2}+0.5)}{\Gamma(m_{r2}+1)} \times\\
  {_2}F_1\left(1,m_{r2}+0.5;m_{r2}+1;\frac{1}{1+c_k}\right), \quad   m_{r2} \ \text{tamsayı değil}
    \end{cases}
    \end{split}
 \end{equation}
olarak bulunur.  Burada $c_k=\frac{\eta_k\rho_r\Omega_{r2}}{2m_{r2}}$, $\rho_r=\sfrac{P_r}{N_0}$ ve $\Omega_{r2}=E[\gamma_{r2}]$ olarak tanımlanır. Ayrıca, $\eta_k=\left(\sqrt{\beta_2}\mp\sqrt{\beta_1}\right)^2$, $k=1,2$ ve $\gamma_{r2}\triangleq\left|h_{r2}\right|^2$ şeklinde verilir.

Benzer şekilde $R-D_1$ arasındaki ortalama hata olasılığı ifadesi (6)-(7) yardımıyla

\begin{equation}
\begin{split}
  &P_1^{(r1)}(e)=\\
  &\begin{cases}
  \sum\limits_{k=1}^5\frac{\tau_k}{2}\left[1-\mu^2(d_k)\sum\limits_{l=0}^{m_{r1}-1}\binom{2l}{l}\left(\frac{1-\mu^2(d_k)}{4}\right)^l\right], \\
  \quad \quad \quad \quad \quad \quad \quad \quad \quad \quad \quad \quad \quad \quad \quad \quad \quad \quad \quad m_{r1} \ \text{tamsayı} \\
  \sum\limits_{k=1}^5\frac{\tau_k}{2\sqrt{\pi}}\frac{\sqrt{d_k}}{\left(1+d_k\right)^{m_{r1}+0.5}}\frac{\Gamma(m_{r1}+0.5)}{\Gamma(m_{r1}+1)} \times\\
  {_2}F_1\left(1,m_{r1}+0.5;m_{r1}+1;\frac{1}{1+d_k}\right), \quad   m_{r1} \ \text{tamsayı değil}
    \end{cases}
    \end{split}
 \end{equation}
olarak bulunur. Burada $d_k=\frac{\zeta_k\rho_r\Omega_{r1}}{2m_{r1}}$, $\Omega_{r1}=E[\gamma_{r1}]$,  \\ $\zeta_k=\left[\left(\sqrt{\beta_2}+\sqrt{\beta_1}\right)^2, \left(\sqrt{\beta_2}-\sqrt{\beta_1}\right)^2, \beta_1, \left(2\sqrt{\beta_2}+\sqrt{\beta_1}\right)^2, \right. \\ \left. \left(2\sqrt{\beta_2}-\sqrt{\beta_1}\right)^2\right]$ ve $\gamma_{r2}\triangleq\left|h_{r2}\right|^2$ şeklinde verilir.

Son olarak $D_1$ kullanıcısı için (7) ve (9), $D_2$ kullanıcısı için ise (5) ve (8) ifadeleri denklem (3)'te yerine konularak e2e hata olasılıkları kapalı formda elde edilir.
\section{Nümer{\footnotesize İ}k Sonuçlar}
Bu bölümde, bir önceki bölümde elde edilen teorik ifadelerin doğrulamaları bilgisayar benzetimleri ile sunulmuştur.  Tüm benzetimlerde, röle gücünün kaynak gücüne eşit olduğu varsayılmıştır ($P_s=P_r$). Bu bölümde sunulan tüm şekillerde, teorik sonuçlar çizgilerle gösterilirken benzetim sonuçları işaretçilerle (marker) sunulmuştur.

Şekil 2'de kullanıcıların hata performansları tüm düğümler arasındaki kanal biçim parametrelerinin eşit olduğu durum için ($m_{sr}=m_{r1}=m_{r2}$) sunulmuştur. Ortalama kanal güçleri $\Omega_{sr}=10dB$, $\Omega_{r1}=10dB$ ve $\Omega_{r2}=0dB$ olarak alınmıştır. Şekil 2'den görüldüğü üzere, benzetim sonuçları elde edilen teorik ifadelerle birebir örtüşmektedir. Ayrıca, biçim parametrelerinin çeşitlilik (diversity) derecesine eşit olduğu açıkça gözükmektedir.
\begin{figure}[!t]
	\centering
	\shorthandoff{=}  
	\includegraphics[width=8.5cm, height=5cm]{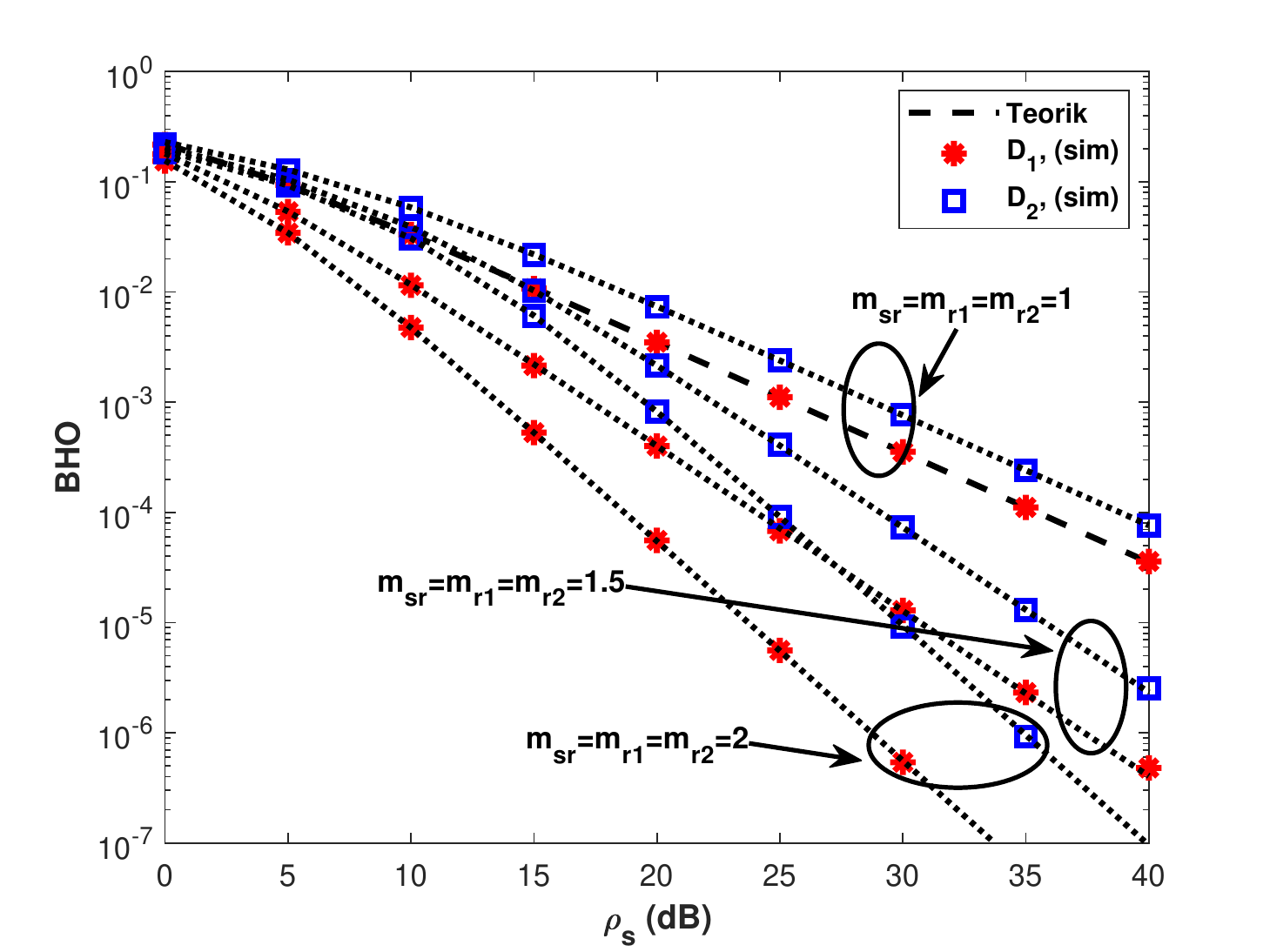}
	\shorthandon{=} 
	\caption{Çöz-aktar işbirlikli NOMA için hata performansı, $\Omega_{sr}=10dB$, $\Omega_{r1}=10dB$ ve $\Omega_{r2}=0dB$ }
	\label{sekil2}
\end{figure}

Şekil 3'te ise düğümlerin biçim parametrelerinin farklı olduğu durumda ($m_{sr}=2$, $m_{r1}=3$ ve $m_{r2}=1$) kullanıcıların hata performansları sunulmuştur. Benzetim sonuçları 3 farklı ortalama kanal güç durumları için verilmiştir. Durum I' de $\Omega_{sr}=3dB$, $\Omega_{r1}=3dB$, $\Omega_{r2}=0dB$, Durum II'de $\Omega_{sr}=6dB$, $\Omega_{r1}=6dB$, $\Omega_{r2}=3dB$ ve Durum II'te $\Omega_{sr}=10dB$, $\Omega_{r1}=6dB$, $\Omega_{r2}=3dB$ olarak varsayılmıştır. Tüm durumlarda açıkca gözükmektedir ki, kullanıcıların uçtan-uca çeşitlilik derecesi düğümlerdeki minimum $m$ parametresi tarafından belirlenmektedir. Bu durum, çöz-aktar protokolünün kullanıldığı işbirlikli iletişim sistemleri için beklenen bir sonuçtur. Şekil 3'te $D_1$ kullanıcısı için çeşitlilik derecesinin $\min\{2,3\}=2$ ve $D_2$ kullancısı için ise $\min\{2,1\}=1$ olduğu kolayca görülebilir. Ortalama kanal güçlerinin değişmesinin ise beklenildiği gibi çeşitlilik derecesine bir etkisi olmamaktadır. Ortalama kanal gücünün artmasıyla kullanıcıların hata performanslarında yatay bir artış gözlenmektedir. Fakat, bu artış biçim parametresi daha düşük olan $D_2$ kullanıcısında daha az miktarda olmaktadır. Hatta, Durum II'den Durum III'e geçişte kaynak-röle arasındaki ortalama kanal güçü artmasına rağmen $D_2$ kullanıcısının hata performansında kayda değer bir değişim gözlenememektedir. Bu durum şu şekilde açıklanabilir: Her ne kadar rölede semboller daha az hatalı çözülse de ikinci zaman diliminde $R-D_2$ arasındaki kanal kalitesinin düşük olması sebebiyle hata oranı artmaktadır ve bu da uçtan-uca hata performansında baskın olmaktadır.
\begin{figure}[!t]
	\centering
	\shorthandoff{=}  
	\includegraphics[width=8.5cm, height=5cm]{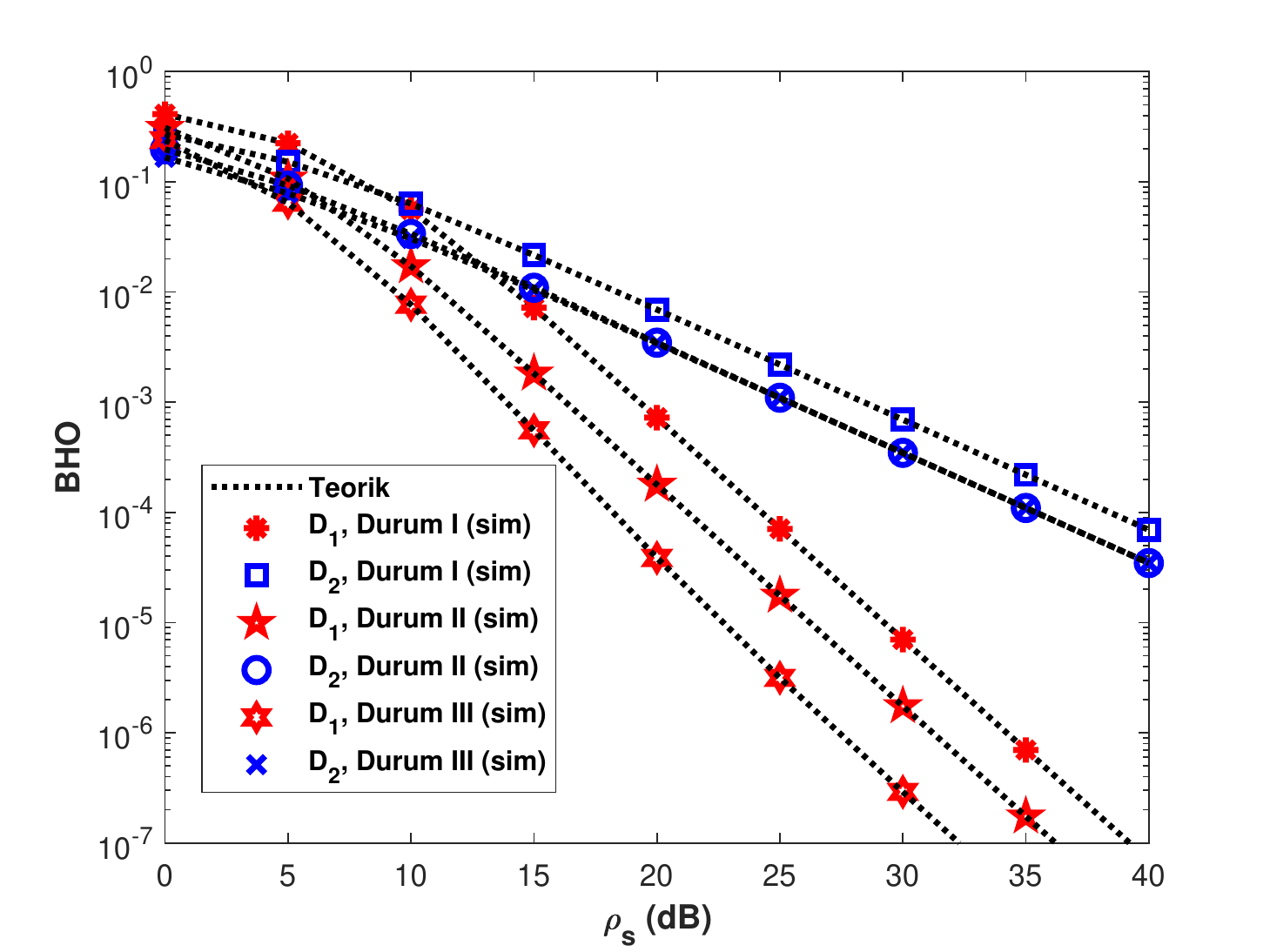}
	\shorthandon{=} 
	\caption{Çöz-aktar işbirlikli NOMA için hata performansı, $m_{sr}=2$, $m_{r1}=3$ ve $m_{r2}=1$}
	\label{sekil3}
\end{figure}

Son olarak kaynaktaki ve röledeki güç paylaşım katsayılarının kullanıcıların hata performansları üzerindeki etkilerini araştırmak adına, güç paylaşım katsayılarının değişimine göre hata performanslarının değişimi Şekil 4'te verilmiştir. Şekil 4'te $\rho_s=20dB$, $m_{sr}=m_{r1}=m_{r2}=2$, $\Omega_{sr}=10dB$, $\Omega_{r1}=10dB$ ve $\Omega_{r2}=0dB$ olarak varsayılmıştır. Beklenildiği üzere kaynakta ya da rölede $D_2$ kullanıcısının sembollerine ($x_2$) aktarılan güç oranı artırıldığında ($\alpha_1$, $\beta_1$ azaldığında), $D_2$'nin hata performansı iyileşmektedir. Fakat, $\alpha_1$ veya $\beta_1$'in artması $D_1$ performansı için her zaman bir kazanç getirmeyebilir. Bu durum şu şekilde açıklanabilir. $x_1$ sembollerine aktarılan gücün çok fazla artması sonucunda, $x_2$ sembollerinin demodüle edilmesi sırasında hata oranı artmaktadır. Hem rölede hem de $D_1$ kullanıcısında SIC işlemi için önce $x_2$ sembolleri demodüle edilmek zorunda olduğu için hata oranındaki bu artış aynı zamanda $D_1$ kullanıcısının hata performansını da düşürmektedir.
\begin{figure}[!t]
	\centering
	\shorthandoff{=}  
	\includegraphics[width=8.5cm, height=8cm]{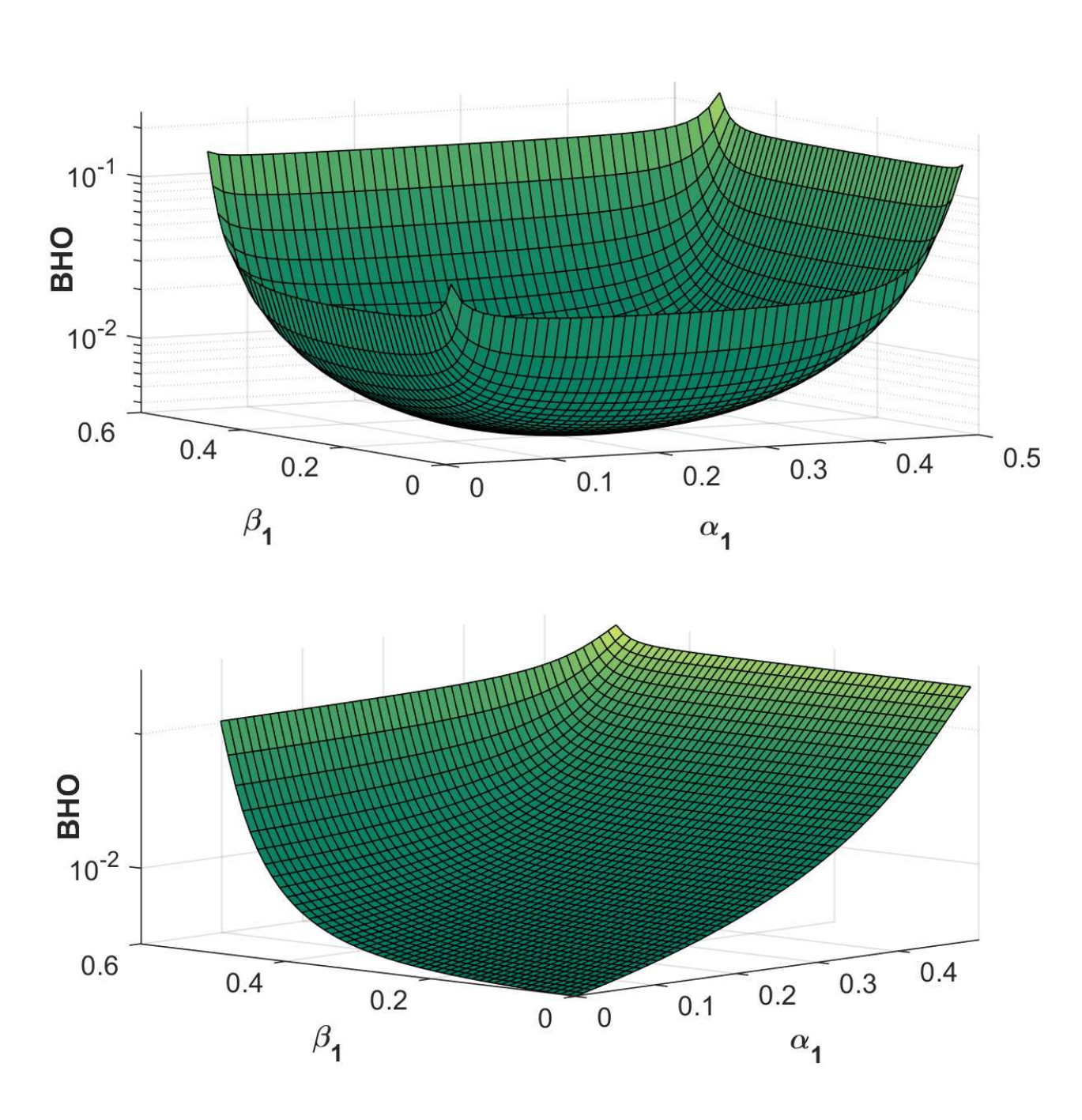}
	\shorthandon{=} 
	\caption{Hata performanslarının güç paylaşım katsayılarına göre değişimi a) $D_1$ b)$D_2$}
	\label{sekil3}
\end{figure}
\section{Sonuç ve Tartışma}
Bu çalışmada çöz-aktar protokolü kullanılan işbirlikli NOMA sistemi için uçtan-uca bit hata olasıkları ifadesi Nakagami-m sönümlemeli kanallarda türetilmiştir. Elde edilen ifadeler bilgisayar benzetimleri desteklenmiştir. Hata performansı üzerinde kanal durumlarının ve güç paylaşım katsayılarının etkileri tartışılmıştır. İşbirlikli NOMA sistemlerinde çeşitlilik derecesinin düğümler arasındaki minimum biçim parametresine ($m$) bağlı olduğu gösterilmiştir. Bu çalışmada sabit güç paylaşım katsayıları kullanılmış olup, farklı kriterler altında optimum güç paylaşımının araştırılması gelecek çalışmalara bırakılmıştır. Ayrıca, bu çalışmanın hem işbirlikli NOMA sistemlerinin farklı kanallardaki hata analizlerine hem de diğer NOMA sistemlerinin hata analizlerine katkı sunacağı beklenmektedir.

\section*{B{\footnotesize İ}lg{\footnotesize İ}lend{\footnotesize İ}rme}

 \textit{Zonguldak Bülent Ecevit Üniversitesi BAP birimi tarafından 2019-75737790-01 numaralı proje ile desteklenmektedir.}




%

\bibliographystyle{IEEEtran}
\bibliography{df_nakag}

\end{document}